\documentstyle[12pt]{article}
\def\la{{\langle}}
\def\ra{{\rangle}}

\newcommand{\beq}{\begin{equation}}
\newcommand{\eeq}{\end{equation}}
\newcommand{\beqa}{\begin{eqnarray}}
\newcommand{\eeqa}{\end{eqnarray}}
\newcommand{\da}{^\dagger}
\newcommand{\wh}{\widehat}
\begin{document}
\baselineskip 14pt

\vspace*{2cm}
\noindent{\Large\bf Quantum arrival time measurement and backflow effect. 
\vspace*{1cm}\\}
{\large J. G. MUGA$^{1,2}$, J. P. PALAO$^1$ and 
C. R. LEAVENS$^2$\vspace{.1cm}\\}
{$^1$ \it Departamento de F\'{\i}sica Fundamental
y Experimental, Universidad de La Laguna,\\
La Laguna, Tenerife, Spain\\}
{$^2$ \it Institute for Microstructural 
Sciences,\\
National Research Council of Canada, Ottawa, Canada
K1A 0R6\vspace*{.3cm}\\}

\noindent{PACS 03.65.Bz Quantum mechanics (foundations)\\
\noindent{PACS 03.65.Nk Nonrelativistic scattering theory \vspace*{.3cm}\\}

\pagestyle{plain}
\noindent{{\bf Abstract}}.--
The current density for a freely evolving state without negative momentum 
components can temporarily be negative. The ``operational arrival time 
distribution'', defined by the absorption rate of an ideal detector, is 
calculated for a model detector and compared with recently proposed
distributions. Counterintuitive features of the backflow regime are
discussed.\vspace*{10cm}\\
%
\newpage
\baselineskip 12pt
The arrival time of a particle at a spatial point is one of the
classical concepts whose quantum counterpart is problematic, even for the
free particle case considered in this paper.
Despite this there are experiments, most
notably ``time of flight
experiments'', that seemingly circumvent the theoretical objections
and difficulties exemplified by Allcock's work [\ref{All}].
Several researchers have tried in recent years to fill this gap between
theory and practice by reexamining the subject using a variety of
approaches [\ref{Busch}-\ref{MSP}]; 
a recent review provides a brief summary of the methods applied
and a discussion of open questions [\ref{MSP}]. 

If all of the particles in an ensemble of freely moving classical particles
have positive momenta $p$ then the distribution $\Pi[t(X)]$ of arrival times
$t(X)$ at the position $x=X$ is just the particle flux
$J[x=X,t]$. (One spatial
dimension is always assumed in this paper, the arrival point $X$ is
taken as the origin $x=0$ unless indicated otherwise, and $t(0)$ is
written simply as $t$.) Not surprisingly,
in many investigations the proposed distribution $\Pi(t)$ of arrival 
times for quantum particles is closely related to the probability
current $J(0,t)$.  In both the quantum and classical cases $J(0,t)$ is
equal to $dN^+(t)/dt$, where $N^+(t)$ is the probability of finding the
particle to the right
of $x=0$ at time $t$ ($N^+(t)=\int_0^\infty |\la x|\psi(t)\ra|^2 \,dx$ 
in the quantum case). Hence, it is tempting to extend
the classical result for the arrival time distribution 
to the quantum regime. But this is not entirely satisfactory
because $J(0,t)$ can be negative for a freely evolving quantum
state even if its momentum distribution is zero for all negative momentum
components, thus invalidating $J(0,t)$ as a probability distribution of
arrival times even in the free particle case.
Bracken and Melloy  showed that 
the time interval over which $J(0,t)<0$ can be arbitrarily
long but finite [\ref{BM}]. They also derived a least-upper-bound, 
estimated to be $0.04$, for the time integral of $|J(0,t)|$ over an
interval of negative $J(0,t)$. But the backflow effect is 
negligible quantitatively at asymptotic distances from
the source or interaction region [\ref{AP95}]. This in part explains
why the arrival time is not particularly worrysome for the practitioner
of time of flight or other arrival time
measurement techniques. However, the fundamental difficulty remains 
and is worth exploring. It is also expected that recent developments
in atomic and optical physics will make the backflow regime amenable
to experimental study.

In this letter we will model an idealized particle detection 
setting and investigate the operational arrival time distribution
focusing on the backflow regime.   
We have in mind a scintillation screen or any other device where 
the detection depends on the passage from the initial channel   
to one or more final channels (associated with changes in chemical
arrangements or internal states of the particle or the apparatus).
It is assumed that the experiment is repeated with single particles
many times,  with the same initial conditions,
and that the
final number of detection counts for a given $dt$ is proportional to
the amount of norm of the initial channel that disappears in that time,
$-dt(dN(t)/dt)$, where $N(t)\equiv\int_{-\infty}^{\infty}dx\,
|\la x|\psi(t)\ra|^2$, 
and $\la x|\psi(t)\ra$ represents the amplitude of the initial channel.
In general, in the absence of backflow at $x=0$, the flux $J(0,t)$
at the front edge of the detector (conventionally located
between 0 and $L$) and $-dN(t)/dt$ are close to
each other, but the latter is slightly delayed 
with respect to the former because of the time it takes to absorb
(i.e., to pass from the incident to the final channels) the 
part of the wave inside the detector. More precisely, 
the time averages evaluated with $J(0,t)$ and $-dN(t)/dt$ differ by the 
mean dwell time in the detector $\tau_D$ [\ref{AP95}].
The arrival time distribution
is in this context defined operationally, it depends on the apparatus,
and it is given by the absorption rate $-dN(t)/dt$ (suitably normalized 
to account for any incident particles that do not reach the detector).
In real detectors additional delays and signal broadening
have to be considered because of the amplification of the microscopic signal,
for example in a photomultiplier, but we shall ignore that 
stage of the process, which is highly dependent on the particular
detection method or apparatus, to concentrate on the generic
first microscopic step.     
     
The effect of such a measuring device on the incident
channel can be characterized by complex reflection and
transmission amplitudes, $R(p)$ and $T(p)$, that depend on the momentum. 
Within the spirit of ``optical models'' a  complex
potential may be constructed subject to the constraint of generating 
functions $R(p)$ and $T(p)$ with specified properties.
This phenomenological approach retains the basic features of
the apparatus-particle system avoiding  the explicit treatment 
of all the degrees of freedom involved.
A good detector will absorb completely over a broad range $\Delta_p$ of
incident momenta, 
\beq
\label{RT}
R(p)=T(p)=0, \;\;\;\;\;\;(p\; {\rm in}\; \Delta_p).
\eeq
With perfect amplification, all particles with incident energies in the
working range will then be detected.
Another desirable feature of a ``time of 
arrival detector'' is that its spatial width $L$
be small. Clearly, these ideal conditions are difficult to meet
in actual detectors, and known complex potentials
do not satisfy them exactly either. It is however
rewarding to study theoretical models as close as possible to the
ideal limit to ascertain what to expect of ``ideal measurements''.  
 
Consider an initial ($t=0$) wavefunction with no negative momentum
components and negligible overlap with the complex potential region $[0,L]$.
In the absence of this potential the freely evolving wavefunction can be
written as [\ref{note}]   
\beq
\la x|\phi_{in}(t)\ra=\int_0^\infty dp \la x|p\ra e^{-ip^2t/(2m\hbar)}
\la p|\phi_{in}(0)\ra\,,
\eeq
whereas, if the potential is present the wavefunction is given by 
\beq\label{po}
\la x|\psi(t)\ra=\int_0^\infty dp \la x|p^+\ra e^{-ip^2t/(2m\hbar)}
\la p|\phi_{in}(0)\ra\,,
\eeq
where the $|p^+\ra$ are scattering eigenstates of $H$ associated with 
the incident plane waves $|p\ra$, and $|\phi_{in}\ra$ is the freely evolving
incoming asymptote to which the state $|\psi\ra$ tends before the collision.  
The validity and meaning of this equation are not trivial. The 
elements of scattering theory with complex potentials required to 
obtain (\ref{po}) are described in the Appendix B.
For a ``perfect absorber'' 
the reflection amplitudes $R(p)$ are zero for those momenta for which 
$\la p|\phi_{in}(0)\ra$ is nonnegligible, 
so for $x\le 0$ the plane wave and the eigenstate of $H$ coincide,
$\la x|p\ra=\la x|p^+\ra$. Therefore,    
$\la x|\psi(t)\ra=\la x|\phi_{in}(t)\ra$ for $x\le 0,$
namely, the wave function to the left of the model detector 
is unaltered by the presence of the perfectly absorbing
complex potential.
The corresponding fluxes for $\phi$ and $\psi$ are
also equal up to the potential edge, 
\beq\label{flu}
J_\psi(x,t)=J_\phi(x,t),\,\;x\le 0\,.
\eeq
This means that it is possible for a perfect absorber to emit probability! 
This counterintuitive phenomenon happens whenever probability backflow occurs 
at $x=0$ in the absence of the complex potential, i.e.  $J_\phi(x=0,t) < 0$
for some finite time range $\delta_t$. Since the perfectly absorbing detector
reproduces this negative flux region exactly it must ``give back'' part of
the probability that had entered $[0,L]$ before $\delta_t$:
$dN^-(t)/dt=-J_\psi(x=0,t) =
-J_\phi(x=0,t) > 0$ for $t$ in $\delta_t$, where
$N^{-}(t)=\int_{-\infty}^0|\la x|\psi(t)\ra|^2\,dx$.
There is, however, no paradox because the leftward probability flow
out of the detector volume is only temporary and does not imply a permanent 
reflection for a fraction of the  
particles since eventually all particles are absorbed.  
Thus a perfect absorber defined according to (\ref{RT}) does not
require that $J(0,t)$ always be nonnegative at its front edge.
Moreover, the fact that $dN(t)/dt\le 0$ at all times for a complex 
potential with a negative imaginary part is not necessarily
inconsistent with $dN^-(t)/dt$ being positive
for a finite range $\delta_t$ of $t$. These surprising, non-classical
results will be demonstrated for an explicit choice of state and complex
potential. The complex potential model is constructed  
by means of a series of complex square barriers with negative imaginary parts,
see  Appendix B. This enables us to know the scattering
eigenfunctions exactly, and to calculate easily
the absorption rate for a given wavepacket by quadrature (also quantities
such as the total absorption, or the dwell time). 
The method
allows for efficient absorption at moderate to large momenta (where 
Eq. (\ref{RT}) can be essentially satisfied with absorptions
higher than $99.9 \%$) but not for very
low momenta. The wave packet is selected accordingly without too low
momenta, and as a consequence the amount of backflow that we shall study
is far from
the upper bound found by Bracken and Melloy [\ref{BM}], but this 
is not essential for demonstrating the effect and 
for illustrating the general theoretical prediction of Eq.
(\ref{flu}).           

We may first consider an  
incident state $|\psi\ra$ that leads to analytical expressions for
$\la x|\psi(t)\ra$
and $J(x,t)$, and that satisfies the stated restrictions of the
arrival time theories of 
references [\ref{Ki},\ref{GRT},
\ref{Gia},\ref{DM},\ref{D}],
\beq\label{mom}
\la p|\psi(0)\ra=C (1-e^{-\alpha p^2/\hbar^2})
e^{-\delta^2(p-p_0)^2/\hbar^2-ipx_0/\hbar}\Theta(p)\,,
\eeq
with $\delta^2 + \alpha > 0$ and $x_0 << -\delta$.
Note the absence of negative
momentum components and the $p^2$ dependence as $p\to 0$.  
The Fourier transform  
can be compactly expressed using 
$w-$functions [\ref{AS}],
\beq\label{wp}
\la x|\psi(t)\ra=C'\left\{
\frac{w(-ig/2A^{1/2})}{A^{1/2}}
-\frac{w[-ig/2(A+\alpha)^{1/2}]}{(A+\alpha)^{1/2}}\right\}\,.
\eeq
Here $w(z)=e^{-z^2}{\rm erfc}(-iz)$, and 
\beqa
A&=&\delta^2+i\hbar t/(2m)\,,\\
g&=&i(x-x_0)+2k_0\delta^2\,,\\
C'&=&\frac{Ch^{1/2}}{4\pi^{1/2}e^{k_0^2\delta^2}}=
\frac{1}{2^{3/2}\pi^{1/4}}
\Bigg\{
\frac{w(-i2^{1/2}k_0\delta)}{2^{3/2}\delta}\,,
\nonumber\\
&-&\frac
{w[-2ik_0\delta^2/(2\delta^2+\alpha)^{1/2}]}
{(2\delta^2+\alpha)^{1/2}}
+\frac
{w\{-2i k_0\delta^2/[2(\delta^2+\alpha)]^{1/2}\}}
{[8(\delta^2+\alpha)]^{1/2}}
\Bigg\}^{-1/2}\,.
\eeqa
where $k_0=p_0/\hbar$.
The derivative with respect to $x$, and therefore the flux,
are also analytical since $dw(z)/dz=-2zw(z)+2i/\pi^{1/2}$.
The momentum distribution of this wave packet is too
close to zero energy to be efficiently 
absorbed, so we shall use instead a new state obtained from (\ref{mom}) 
by a  ``boost'' at time $t=0$,
\beq\label{boost}
\la p|\psi'(0)\ra=\la p-b|\psi(0)\ra\,.
\eeq
The flux for the new state can be related to the flux and 
probability density of the original one,
\beq\label{fluxes}
J_{\psi'}(0,t)=
J_{\psi}(-bm/t,t)+(b/m)|\la x=-bt/m|\psi(t)\ra|^2\,. 
\eeq
The last term may be understood as the flux contribution due to
the displacement of an observer with velocity $-b/m$. Since 
$b>0$, it is always positive, 
so if the boost is strong enough the backflow region
disappears. We have chosen a value of $b$ small enough that the effect
remains and large enough that essentially full absorption can be achieved
(99.97\%).
The complex potential has been improved in successive optimizations until 
the prediction of Eq. (\ref{flu}) has been numerically confirmed by the 
converged flux results.        
Figures 1 and 2 show the flux at $x=0$ (with and without complex potential),
and $-dN(t)/dt$. The two fluxes, with and without complex
potential, are  indistinguishable on the scale of the figures. This means 
in particular that, during the backflow regime detailed in Figure 2, the
potential region $[0,L]$ returns probability to the 
left half space $[-\infty,0]$ at exactly the rate required to maintain the
same negative flux as the freely evolving state. Fig. 1 shows global 
agreement in shape between the flux and the absorption rate
$-dN/dt$, the latter being slightly delayed.
It is of interest to compare the 
operational arrival time distribution $-dN(t)/dt$ with other 
theoretical proposals for the time of arrival distribution. 
Also shown in Figures 1 and 2 are Kijowski's distribution 
\beq\label{Pi}
\Pi_K(t)=\left|\frac{1}{(mh)^{1/2}}\int_0^\infty \sqrt{p}\,
e^{-ip^2t/2m\hbar}
\la p|\psi(0)\ra\,dp\right|^2\,, 
\eeq
and that derived 
from Bohm's causal theory [\ref{deriv},\ref{com}]
\beqa 
\Pi_B(t)=|J(0,t)|/\int_0^\infty |J(0,t')| dt', 
\eeqa
both evaluated for $\psi'$. The former distribution
was originally derived by imposing a series of conditions consistent 
with the classical distribution [\ref{Ki}], 
and has been later studied, rederived or generalized by several authors
[\ref{Busch},\ref{GRT}-\ref{D}] [The transformation
(\ref{fluxes}) does not hold when $J$ is replaced by $\Pi_K$ because of the 
non linearity introduced by the square root in (\ref{Pi})].
It arises naturally as the square of
the overlap between the initial state (restricted to positive momenta) 
and the eigenstates of the ``time of arrival 
operator'' 
\beq       
\widehat{t}=
-\frac{m}{2}\left(\widehat{q}
\frac{1}{\widehat{p}}+\frac{1}{\widehat{p}}
\widehat{q}\right)\,.
\eeq
[The different quantization 
in [\ref{GRT}] gives the same result.]
As for any other quantity obtained from a formal quantization procedure,
its physical meaning and content is not immediately obvious and
requires detailed examination and external justification, 
since in principle other quantizations with the same classical
limit may also be constructed [\ref{Leon},\ref{MSP}]. 
In the numerical example Kijowski's distribution $\Pi_K(t)$ is in overall
excellent agreement with the flux (except at the fine scale needed to 
resolve the backflow region) and, up to the delay, with $-dN(t)/dt$.
With the latter it has in common positivity at all times. The two quantities
avoid the negative values of the backflow region smoothly while
$|J(0,t)|$ has downward cusps at the zeroes of $J(0,t)$.
In Bohm's theory 
particle trajectories do not intersect each other so that only a single
trajectory contributes to the current density $J(x,t)$ at each space-time
point $(x,t)$. For times $t$ before $\delta_t$, when $J(0,t)
> 0$, particles 
in the ensemble crossing $x=0$ do so only from left to right; for times $t$ 
during $\delta_t$, when $J(0,t)<0$, some of these particles recross
$x=0$, moving from right to left this time - there are no particles crossing
$x=0$ from left to right during $\delta_t$; for times $t$ after $\delta_t$,
when $J(0,t)>0$ again, particles cross $x=0$ only from left to right.
By continuity, at those instants of time $t$ when $J(0,t)=0$ no particle in 
the ensemble is crossing $x=0$ from either direction, leading to the
above mentionned cusps. From a classical 
point of view, the counterintuitive
aspect of this picture of backflow, when applied to an ensemble of freely
evolving particles, is the necessity for some of these ``free'' particles
to (twice) come to rest and then reverse their direction of motion. However 
within Bohm's theory the particles are not truly free but
are guided by the
wave function $\psi(x,t)$ [\ref{Bohm}] which itself undergoes
relatively rapid temporal
changes in the backflow region considered here. Kijowski's distribution  
also contains an obvious counterintuitive feature and is subject to
an interpretational puzzle. In different 
publications [\ref{Ki},\ref{GRT},\ref{DM},\ref{D}], 
which all lead to the arrival time distribution (\ref{Pi}) for a
freely evolving
state containing no negative momentum components, it is claimed that under
those circumstances particles arrive at $x=0$ only from the left. In
the backflow regime this leads to the baffling conclusion that
the probability of finding the particle to the left of $x=0$ is increasing
with time in a time interval during which particles reach $x=0$
only from the left. According to Bohm's theory, the particles that
contribute to the backflow
region shown in Fig. 2  arrive at $x=0$ three times, twice from the left
and once from the right, so $\Pi_B(t)$ is not a ``first arrival'' time
distribution; according to the approaches leading to Kijowski's
distribution (\ref{Pi}) they arrive only from the left but it is not
obvious how this is consistent with $dN^-(t)/dt >0$  and whether
or not one should regard (\ref{Pi}) as a distribution of first
arrival times. Finally, the operational distribution $-dN/dt>0$ has the
advantage of a clear physical content, and direct relation to an
experimental setting, but it should not be
overinterpreted, in particular since the present analysis has shown that 
a perfect absorber can emit probability.
\section*{Appendix A: Construction of complex absorbing potential}
Let us find a potential with support $[0,L]$ that maximizes the 
absorption in a given momentum interval $[p_1,p_2]$.
The method described here is more efficient and
numerically robust than previous ones [\ref{LL}].
The proposed functional form is a series of equal length $N$
complex square barriers
with complex energies 
$\{V_j\}, \,j=1,2,...,N$. The real and imaginary values of $V_j$
are found by minimizing  
(with the restriction Im$(V_j)<0$),  
the sum of the survival probabilities at $s$
values of $p$, $\{p_\alpha \}$,
\begin{equation}\label{f}
f(V_1,...,V_N;p_1,...,p_s)
=\sum_{\alpha=1}^{s}S(V_1,...,V_N;p_\alpha)\,,
\end{equation}
where $S(p)\equiv 1-|T(p)|^2-|R(p)|^2$, and  
the $s$ points are evenly spaced in the  
absorption interval $[p_1,p_2]$.
$S$ and its gradient with respect to $\{V_j\}$ are  obtained
by multiplication of known $2\times 2$ transfer matrices  
so that the optimizations are very fast.  
For the application in the main text we have taken $L=0.01$,
$N=4$,  $s=49$, $p_1=260$, and $p_2=740$. 
\section*{Appendix B: Elements of complex potential scattering
theory in one dimension}
For the formulation of scattering theory of complex non hermitian
potentials it is necessary to consider scattering eigenstates of 
$H$, $|p^+\ra$, and their biorthogonal partners $|\wh{p}^+\ra$,
which are eigenstates of 
$H\da$, 
\beqa
|p^+\ra&\equiv&|p\ra+\frac{1}{E_p+i0-H}V|p\ra\\
|\wh{p}^+\ra&\equiv&|p\ra+\frac{1}{E_p+i0-H\da}V\da|p\ra\,.
\eeqa
M\"oller operators that connect the actual state $\psi$ that evolves,
respectively, 
with $H$ or $H\da$, with the freely
evolving incoming asymptotic
states (to which $\psi$ tends before the collision) can be 
defined as $\Omega_+=\int dp\, |p^+\ra\la p|$ and 
$\wh{\Omega}_+=\int dp\, |\wh{p}^+\ra\la p|$ respectively.
They obey  
\beqa\label{gir}
\wh{\Omega}_+\da \Omega_+&=&1_{\rm op}\,,
\\
{\Omega}_+\wh{\Omega}_+\da&=&1_{\rm op}-\Lambda\,,
\eeqa
where $\Lambda=\sum_j |\Psi_j\ra \la\widehat{\Psi}_j|$ is given in terms
of the bound states of $H$ ($\Psi_j$), and of $H\da$, ($\widehat{\Psi_j}$).  
Thus the
integral form of an arbitrary  wave packet evolving with $H$ without 
bound state component can be written as  
\beq\label{wpa}
\la x|\psi(t)\ra=\int dp\,\la x|p^+\ra
e^{-iE_pt/\hbar}\la \wh{p}^+|\psi(0)\ra\,.
\eeq
Using the generalized isommetry relation
(\ref{gir}) one may
substitute $\la\wh{p}^+|\psi(0)\ra$ by $\la p|\phi_{in}(0)\ra$
in (\ref{wpa}).
Moreover, if $\psi(0)$ does not significantly overlap
with the potential and there are not negative momenta, $\phi_{in}(0)$
can also by substituted by $\psi(0)$ as shown in [\ref{MBS}].    
\newpage
\centerline{\bf REFERERENCES\vspace*{.2cm}\\}
\begin{enumerate}
\item\label{All} ALLCOCK G. R., {\it Ann. Phys.} (NY) {\bf 53} (1969) 253;
{\bf 53} (1969) 286; {\bf 53} (1969) 311.
\item\label{Busch} BUSCH P., GRABOWSKI M., LAHTI P. J., 
{\it Phys. Lett.} A {\bf 191} (1994)357.
\item\label{Lea} McKINNON R. and LEAVENS C. R., {\it Phys. Rev. A}
{\bf 51} (1995) 2748.
\item\label{AP95} MUGA J. G., BROUARD S. and MACIAS D., {\it Annals
of Physics} (NY) {\bf 240} (1995) 351.
\item\label{GRT} GROT N., ROVELLI C. and TATE R. S., {\it Phys. Rev.}
A {\bf 54} (1996) 4676.
\item\label{Daumer} DAUMER M., in {\it Bohmian Mechanics and Quantum
Theory: An Appraisal}, edited by J.T. CUSHING, A. FINE and S. GOLDSTEIN,
p. 87
(Kluwer Academic, Dordrecht) 1996.
\item\label{Ar} JADCZYK, A. and BLANCHARD, PH, {\it Helv. Phys. Acta}
{\bf 69}, 613 (1996). 
\item\label{Leon} LEON J., {\it J. Phys.} A  {\bf 30} (1997) 4791. 
\item\label{Gia} GIANNITRAPANI R., {\it Int. J. Theor. Phys.} {\bf 36}
(1997) 1575.
\item\label{DM} DELGADO V. and MUGA J. G., {\it Phys. Rev.} A {\bf 56}
(1997) 3425.
\item\label{Aha} AHARONOV Y., OPPENHEIM J., POPESCU S., 
REZNIK B., UNRUH W. G., {\it Phys. Rev.} A, to appear.
\item\label{D} DELGADO V., {\it Phys. Rev.} A, to appear.
\item\label{Ha} HALLIWELL J. J. and ZAFIRIS E., {\it Phys. Rev. D},
to appear
\item\label{MSP} MUGA J. G., SALA  R. and PALAO J. P.,
{\it  Superlattices and 
Micrstuctures}, to appear.
\item\label{BM} BRACKEN A.J. and MELLOY G.F., {\it J. Phys. A: Math. Gen.}
{\bf  27} (1994) 2197.
\item\label{note} A wavepacket with no negative 
momenta has always spatial support on the full real line, but 
the norm on the right $N^+$ can be made arbitrarily small.
\item\label{Ki} KIJOWSKI J., {\it Rep. Math. Phys.} {\bf 6} (1974) 362.
\item\label{AS} ABRAMOWITZ M. and STEGUN I. A., {\it Handbook of
Mathematical Functions} (Dover, New York) 1972.
\item\label{deriv} LEAVENS C. R., in {\it Tunnelling and its
Implications}, edited by D. MUGNAI,
A. RANFAGNI and L.S. SCHULMAN (World Scientific, Singapore) 1997, p. 100.
\item\label{com} This result also applies in the presence of a
nonzero potential and for an incident state containing both positive and
negative nonzero momentum components. For the special case considered
in this paper, the
denominator can be replaced by unity.
\item\label{Bohm} BOHM D. and HILEY B. J., {\it The Undivided Universe}, 
(Routledge, London) 1993; 
HOLLAND P. R., {\it  The quantum theory of motion} (Cambridge
Univ. Press, Cambridge) 1993,
and references therein.
\item\label{LL} BROUARD S., MACIAS D. and MUGA J. G.,
{\it J. Phys. A} {\bf 27}, (1994) L439.
\item\label{MBS} MUGA J. G., BROUARD S. and SNIDER R. F.,
{\it Phys. Rev.} A {\bf 46} (1992) 6075. 
\end{enumerate}

\newpage
{\large FIGURE CAPTIONS\vspace*{.3cm}\\}

{\bf Figure 1:} $J(0,t)$ for free motion (solid line),
$J(0,t)$ with the absorber, (short dashed line),
$\Pi_K(t)$ (dotted-dashed),
and $-dN/dt$ (long dashed line) for the wave packet 
$\psi'$ in (\ref{boost}), see also (\ref{wp}), with the following
parameters (all in atomic units): 
$\alpha=1.4$, $p_0=1$, $x_0=-0.22$, $\delta=0.007$, $b=300$.
The first three curves are indistinguishable in this scale.   
   
{\bf Figure 2:}  $J(0,t)$ for free motion (solid line),
$J(0,t)$ with the absorber (short dashed line), $|J(0,t)|$ (dotted line),
$\Pi_K(t)$ (dotted-dashed line), $-dN(t)/dt$ (dashed line),
and $-dN(t)/dt|_{y+\tau_D}$ (dashed line with squares). Same parameters as 
in Figure 1. The first two curves are hardly distinguishable. 
$\tau_D=1.0515\times 10^{-5} au$.     
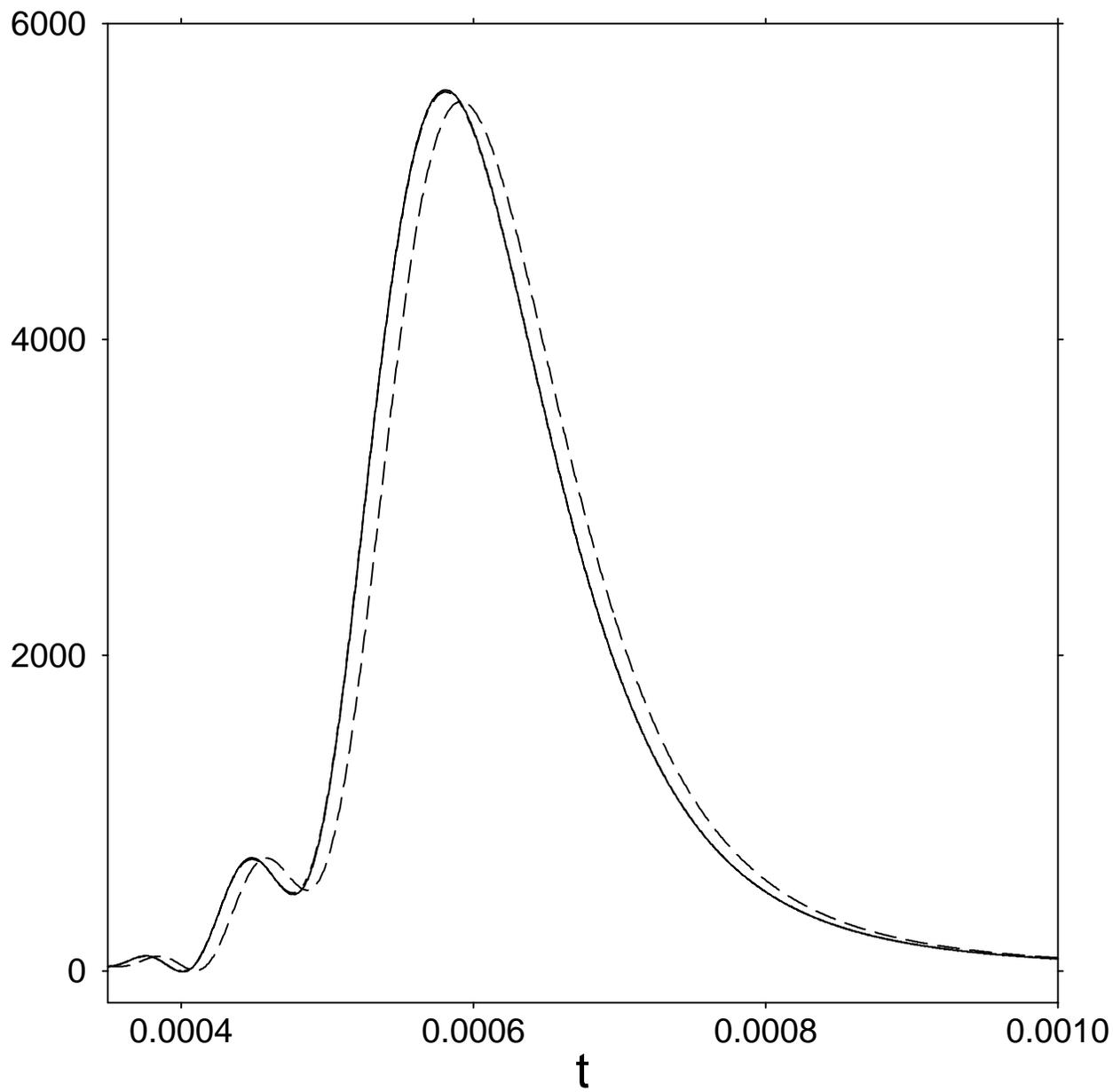
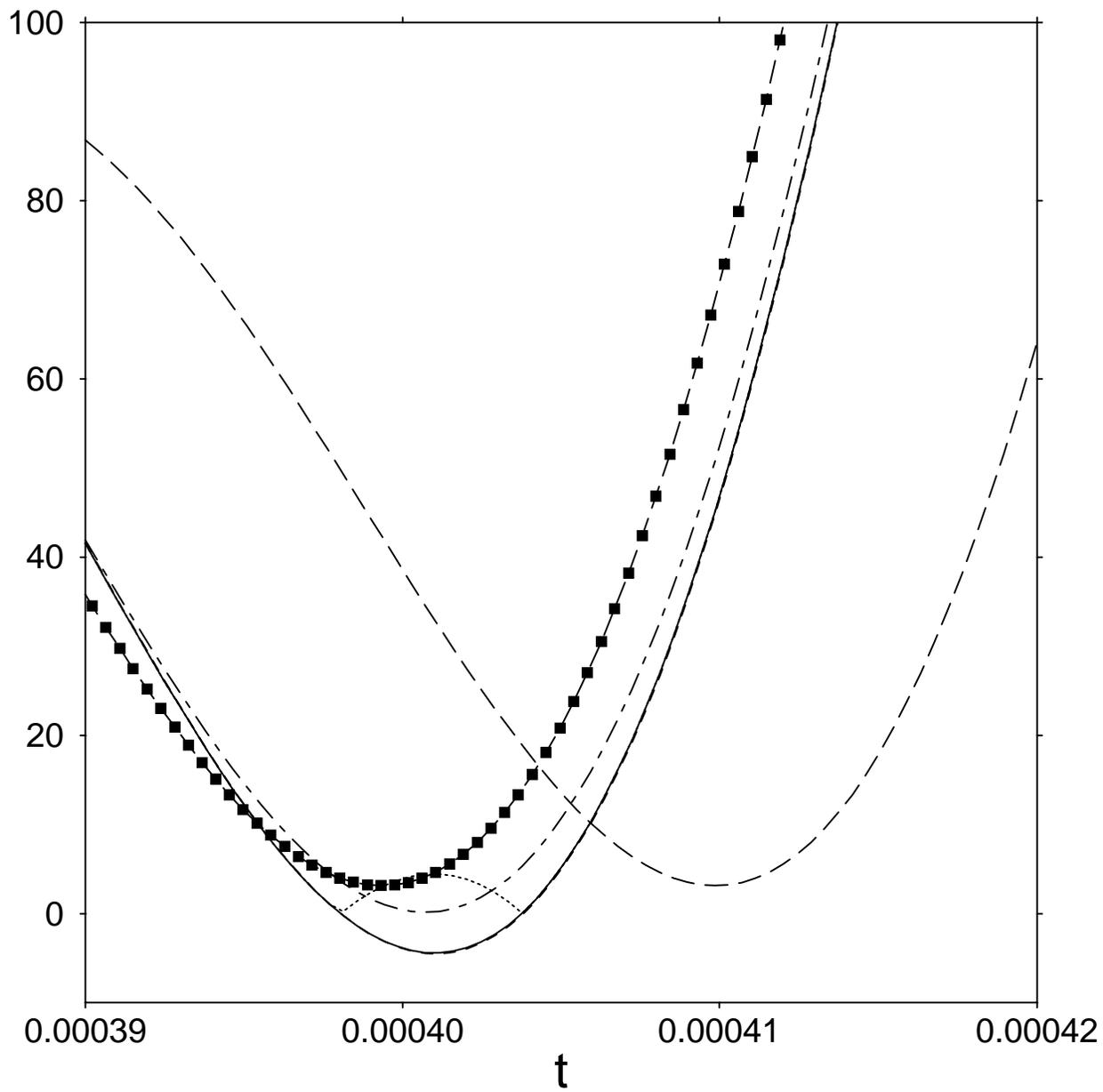
\end{document}